\documentclass[prb,twocolumn,aps,showpacs,superscriptaddress]{revtex4}
\usepackage{color}
\usepackage{amsmath}
\usepackage{graphicx}

\begin{document}

\title{A dynamical mean-field approximation study of a tight-binding model for Ga$_{1-x}$Mn$_{x}$As} 

\author{ A.-M. Nili}
\address{Department of Physics and Astronomy \& Center for Computation and 
Technology, Louisiana State University, Baton Rouge, Louisiana 70803, USA}
\author{U. Yu}
\address{Department of Applied Physics, Gwangju Institute of Science and 
Technology, Gwangju 500-712, Korea}
\author{J. Moreno}
\author{D. Browne}
\author{M. Jarrell}
\address{Department of Physics and Astronomy \& Center for Computation and 
Technology, Louisiana State University, Baton Rouge, Louisiana 70803, USA}

\date{\today}

\begin{abstract}
The magneto-optical properties of the ferromagnetic semiconductor Ga$_{1-x}$Mn$_{x}$As are studied
within the dynamical mean-field approximation (DMFA). A material-specific multiband $sp^{3}$ 
tight-binding Hamiltonian is employed for the dispersion of the GaAs host. The calculated density 
of states shows an impurity band and a distorted valence band for large and moderate values of magnetic 
coupling, respectively. Upon using the more realistic band structure, the ferromagnetic transition 
temperature is significantly closer to the experimental results than the previous predictions of 
$k\cdot p$ models. The optical conductivity shows a Drude-like peak at low frequencies which is 
suppressed by increasing of the magnetic coupling.
\end{abstract}


\maketitle

\paragraph*{Introduction-}  
The discovery of a new generation of dilute magnetic semiconductors (DMS) with large Curie temperature 
\cite{h_ohno_98,f_matsukura_98} has led to numerous experimental and theoretical 
studies.\cite{t_dietl_00,t_jungwirth_06} The ultimate goal is to find a DMS with a ferromagnetic 
transition above room temperature and suitable for use in 
spintronic devices. One of the most promising candidates is 
Ga$_{1-x}$Mn$_{x}$As due to its rather high $T_{c}$, \cite{t_dietl_00,h_ohno_98} and compatibility
with current electronic applications.
Despite this interest, some of the most basic questions about Ga$_{1-x}$Mn${_x}$As are still 
unanswered after more than a decade of its discovery. Among them are questions about  the 
nature of its underlying magnetic interactions and the role of the impurity band. 

In this work we combine the $sp^{3}$ 
tight-binding Hamiltonian for the zinc blende GaAs with the self consistent 
dynamical mean-field approximation (DMFA) to study the magnetic and transport properties of 
Ga$_{1-x}$Mn$_{x}$As. 
We assume that the manganese (Mn) doping does not alter the tight-binding parameters of GaAs and any change 
in the band structure is due to the many-body effects induced by the magnetic coupling
between the Mn local moments and the itinerant holes. 

Several previous studies have used the tight-binding approximation for DMS. 
Tang and Flatt\'e calculated the local density of states for a single Mn and two nearby Mn 
impurities\cite{j_tang_04} and the magnetic circular dichroism\cite{j_tang_08} of bulk Ga$_{1-x}$Mn$_{x}$As 
using large supercells; Ma\v{s}ek \textit{et al.} took the Weiss mean-field approach to calculate the electronic 
structure for several DMS\cite{j_masek_07a}; and  Turek \textit{et al.}
compared both tight-binding approaches\cite{m_turek_08,m_turek_09}. Large-scale 
Monte-Carlo studies of real-space tight-binding Hamiltonians including only the three valence bands 
has also been performed~\cite{y_yildirim_07}.

Our DMFA calculation incorporates quantum self-energy corrections which were
not included in most of the studies described above. We also incorporate valence and conduction bands 
on an equal footing while Yildirim {\it et al.}~\cite{y_yildirim_07} included only valence bands.   
Because this method is non-perturbative, it allows us to study both the metallic
and impurity-band regimes, as well as small and large couplings.
Since the strength of the coupling between the magnetic ions and the charge carriers is comparable 
to the Fermi energy, temporal fluctuations included in our DMFA approach are required
in a realistic calculation especially in the vicinity of the critical temperature.    

We choose tight-binding parameters according to Chadi and Cohen~\cite{d_chadi_75,d_chadi_77}, which give the 
correct band features within a sufficiently wide window of $\pm$~2.0~eV around the Fermi energy, such as 
effective masses for valence and conduction bands and the gap at the center of the Brillouin zone.
Our model also includes the spin-orbit interaction of the parent material which was proved very important in  
previous studies of Ga$_{1-x}$Mn$_{x}$As.~\cite{g_zarand_02a,j_moreno_06a,u_yu_10,a_nili_10a} 
The inclusion of the realistic band structure of the parent material leads to more realistic 
results. For example, the band repulsion between the conduction band and the impurity band, which was absent in 
some previous calculations based on the $k \cdot p$ 
approach~\cite{k_aryanpour_05a,u_yu_10,a_nili_10a,m_majidi_06a},
results in a significant reduction of the bandwidth of the impurity band. Narrowing of the impurity bandwidth  
increases the localization of the mediating 
holes, which suppresses the estimated critical temperature. Consequently, we find that the calculated 
$T_{c}$ is significantly smaller, and closer to the experimental results.

\paragraph*{Model-}

We model the magnetic interactions in Ga$_{1-x}$Mn$_{x}$As using the simplified Hamiltonian proposed by 
$\rm{Zar\acute{a}nd}$ and $\rm{Jank\acute{o}}$:\cite{g_zarand_02a}
\begin{equation}
H= H_{0} + J_{c} \sum_{i} \mathbf{S}(R_i) \cdot \mathbf{J}(R_i) ,
\label{hamiltonian}
\end{equation}
where $H_{0}$ is a 16$\times$16 matrix including both electronic dispersion and spin-orbit coupling of the 
$sp^{3}$ holes of the parent material. Our tight-binding parameters reproduce 
the correct band structure for the heavy and light bands around the center of the Brillouin zone, 
the split-off energy gap and the gap between the valence and the conduction 
bands\cite{d_chadi_75,d_chadi_77}. 
The spin-orbit coupling is modeled with a  term $\lambda_{\alpha} {\bf L} \cdot {\bf S}$ 
where $\lambda_{\rm Ga}$ and $\lambda_{\rm As}$ are interaction constants for Ga and As atoms respectively.
The second term in Eq.~(\ref{hamiltonian}) describes the interaction 
between Mn spins and mediating holes, where $ J_{c}$ 
is the exchange coupling between the localized moments with spin $\mathbf{S}(R_i)$
and the holes total angular momentum density  
$\mathbf{J}(R_i)$, both at site $R_i$. Since the spins of 
Mn$^{+2}$'s are relatively large ($S=5/2$) we treat them as classical vectors. 

\begin{figure}[t]
\begin{centering}
\includegraphics[width=0.45\textwidth]{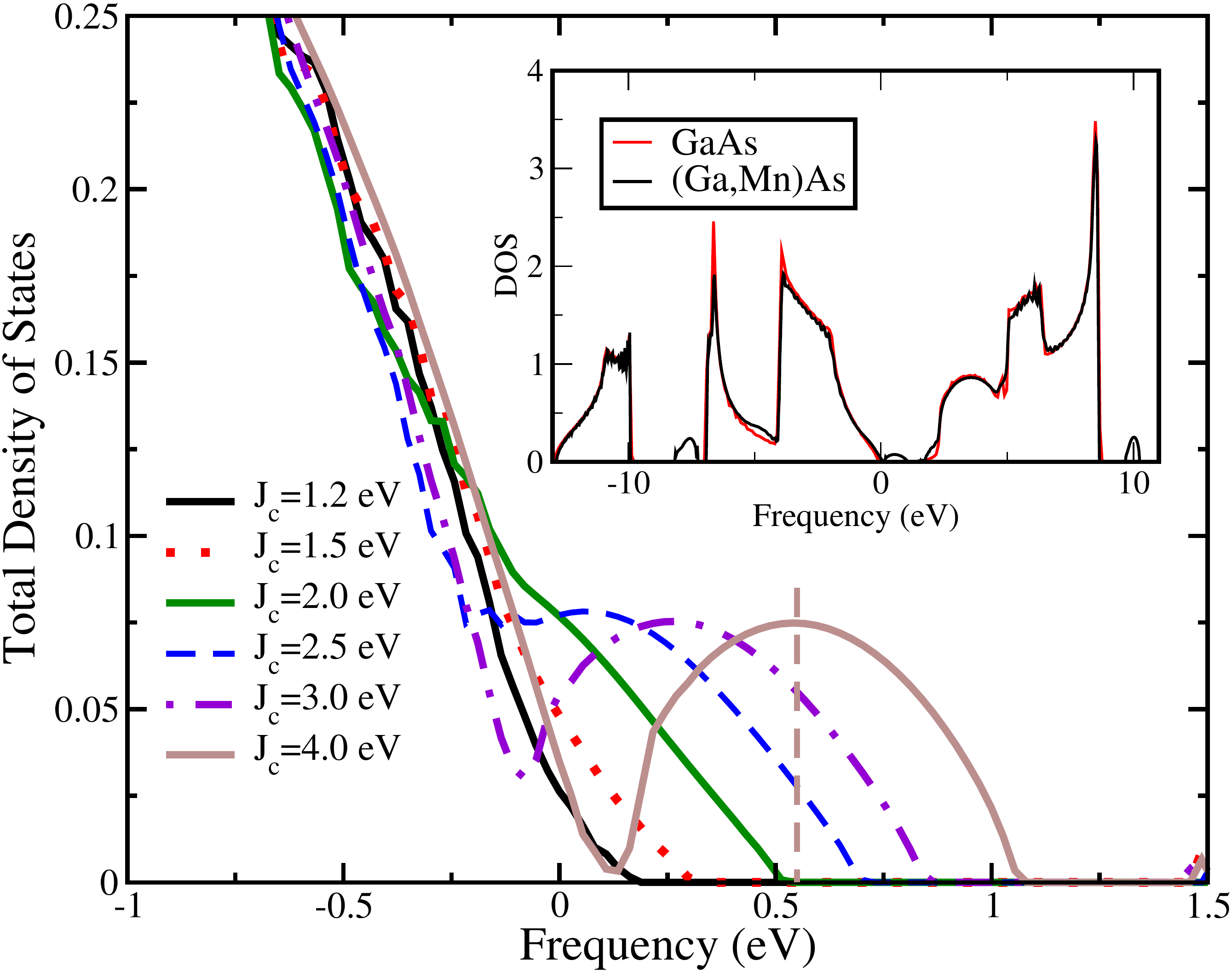}
\caption {(color online). Total density of states versus energy at the edge of the valence band for 
$x=0.05$, $n_{h}=x/2=0.025$ and several coupling strengths at $T$=58~K. The edge of the conduction band 
lies at the right bottom corner. The impurity band is not well separated from the 
valence band up to large couplings ($J_{c}\sim$ 4~eV). The vertical dashed line shows the location of the 
chemical potential for $J_{c}$=4.0~eV. Band repulsion 
between the conduction and impurity bands leads to reduction of the impurity bandwidth.
Inset:  $sp^{3}$ total density of states for pure GaAs and Ga$_{0.95}$Mn$_{0.05}$As with $J_{c}$=4.0 eV. 
We use a Fourier 
transform filter to reduce the noise in the calculated density of states. The noise is in regions 
away from the Fermi energy and is due to the effect of the $\mathbf k$ mesh in the 
coarse graining step of the DMFA self-consistency loop.
We have used up to 62,000 $\mathbf k$ points in this calculation. }
\label{dos_edge}
\end{centering}
\end{figure}

We employ the DMFA algorithm~\cite{a_georges_96a}
to calculate the magnetic and transport properties of the material within a Green's function 
formalism~\cite{k_aryanpour_05a,m_majidi_06a}.
We calculate the average magnetization of the manganese ions: 
\begin{equation}
M=\frac{1}{{\cal Z}} \int d\Omega_{s} S^{z} \exp\{-S_{\rm eff}(\mathbf S)\}
\end{equation}
with the partition function
${\cal Z} = \int d\Omega_{s} \text {exp[-S}_{\rm eff}(\mathbf S)]$,
and the effective action  
$S_{\rm eff}(\mathbf S)=-\sum_{n} \log\det[\hat{\cal G}_{0}^{-1}(i\omega_{n}) +
J_{c} {\mathbf S} \cdot \hat{\mathbf J}]$,~\cite{n_furukawa_98a,n_furukawa_94a} 
where $\hat{\cal G}_{0}$ is the cluster excluded Green function.

The average polarization of the charge carriers at the chemical potential is defined as: 
\begin{equation}
{\cal P}=\frac{\sum_{i}j^{z}_{i}n_{i}(\mu)}{\sum_{i}n_{i}(\mu)}
\label{polar_formula}
\end{equation} 
where $j^{z}_{i}$ is the $z$ component of the total angular momentum and
$n_{i}$, the density of holes in the $i$th band. 

To compute the optical conductivity we use the Kubo formula:\cite{r_kubo_57}

\begin{equation}
\begin{aligned}
\sigma_{ij}(\omega)=\frac{\pi}{\omega} \int d\nu [f(\nu)-f(\nu+ \omega)] \times \\
 \sum_{\rm \textbf{k}} Tr[v_{i}({\rm \textbf{k}})A({\rm \textbf{k}},\nu+\omega)v_{j}({\rm \textbf{k}})
A({\rm \textbf{k}},\nu)],
\end{aligned}
\label{optical_cond}
\end{equation}
where $Tr$ is the trace, $v_{i}(\mathbf k)$ is the velocity along the $i$ direction,
$A(\mathbf k,\omega)$, the spectral function at momentum $\mathbf k$ and frequency $\omega$, and 
$f(\omega)$, the Fermi function. The group velocities involve the hopping terms in the Hamiltonian 
and are calculated as 
$\displaystyle v_{i}(\mathbf k)=\frac{\delta H(\mathbf k)}{\delta A_{i}}  |_{A=0}$, where $H$ 
is the Hamiltonian and $\mathbf A$ the vector potential. 

\paragraph*{Results and Discussion-}

\begin{figure}[t]
\begin{centering}
\includegraphics[width=0.45\textwidth]{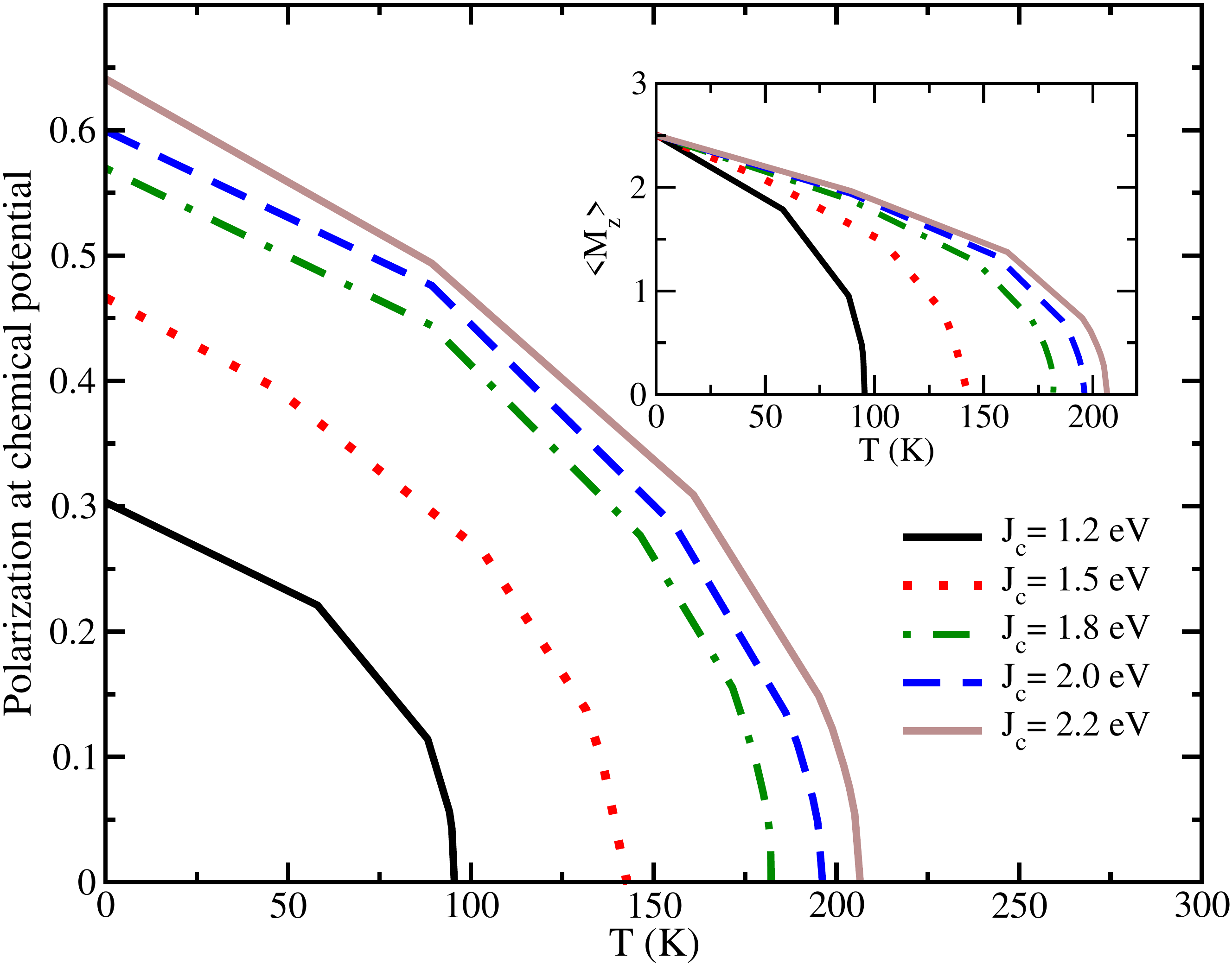}
\caption{(color online). Average hole polarization versus temperature for different values of $J_{c}$ at 
$n_{h}=x/2=0.025$. The polarization is frustrated since it does not reach its maximum allowed value, 
${\cal P}=3/2$, at $T=0$. Comparison with Figure \ref{dos_edge} shows that noticeable changes 
in $T_{c}$  happens when the impurity band is forming. 
Inset: Average magnetization of the ions versus temperature for the same couplings.}
\label{pol.vs.t}
\end{centering}
\end{figure}
We first focus on the density of states (DOS) of the charge carriers at the doping $x$=0.05 
where $T_{c}$ is among the highest reported.\cite{h_ohno_98,d_chiba_03,h_munekata_89} In Figure
\ref{dos_edge} and \ref{pol.vs.t} we assume 50\% hole compensation due to 
anti-site and interstitial doping, such that  n$_{h}=x/2=0.025$. This corresponds to 
the optimum filling in the impurity band regime. In any case, the profile of the DOS changes 
minimally with filling. Figure~\ref{dos_edge} shows the total 
density of states at the edge of the valence band for several values of the coupling. 
Most experimental probes, such as photoemission \cite{j_okabayashi_98}, infrared 
\cite{m_linnarsson_97,a_bhattacharjee_99,e_singley_03} and resonant
tunneling \cite{h_ohno_98b} spectroscopies, and magneto-transport experiments \cite{t_omiya_00}
infer a value of $J_c$ between 0.6-1.5 eV. Within this range of couplings 
the effect of the magnetic interactions is just a distortion at the edge of the valence band. We have to use 
coupling strengths as large as 2.5 eV to observe the formation of the impurity band.
However, recent scanning tunneling microscopy experiments~\cite{a_richardella_10} 
display an impurity band similar to the one appearing in 
our results for couplings around $J_{c}\sim$~3.0 eV. The impurity band is not completely separated until 
couplings  around $\sim$ 4 eV. 
As we expected, the repulsion between the conduction and the impurity band confines the latter within the 
band gap and narrows down the impurity bandwidth. This leads to localization of the  charge 
carriers which in return reduces the ferromagnetic transition temperature, $T_{c}$.

\begin{figure}[t]
\begin{centering}
\includegraphics[width=0.45\textwidth]{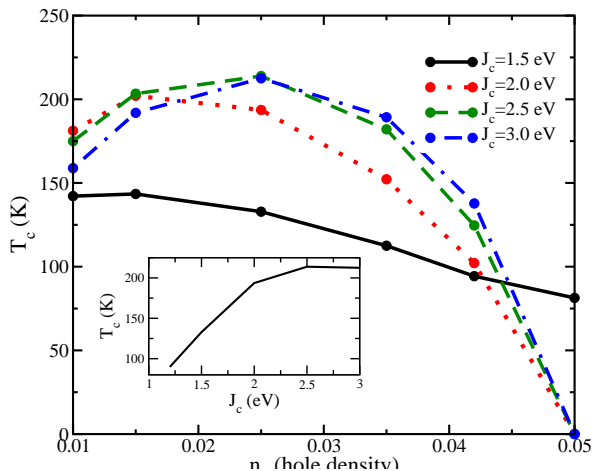}
\caption{(color online). $T_{c}$ versus filling at constant doping, $x$=5$\%$, for different couplings.
For small couplings $T_{c}$ decreases by increasing the hole concentration. For larger
couplings there is an optimum filling of $n_{h}$=2.5$\%$ where $T_{c}$ is maximum.
For coupling values lager than 2 eV,  $T_{c}$ does not change appreciably. The inset shows 
$T_{c}$ vs. coupling strength for $n_{h}=x/2$. The saturation of $T_{c}$
is due to the lack of non-local correlations in the DMFA.}
\label{Tc.vs.nf}
\end{centering}
\end{figure}

Figure~\ref{pol.vs.t} displays the temperature dependence of the hole polarization
for different couplings. We choose a narrow energy window ($\sim$0.1 eV) at the chemical potential 
and calculate the average ${\cal P}$ within this window using Eq.~(\ref{polar_formula}). 
The polarization is frustrated since it does not reach its maximum allowed value, 
${\cal P}=3/2$, at $T=0$, and it is further reduced with 
decreasing magnetic coupling strength. Comparison with Figure~\ref{dos_edge} shows that  noticeable 
changes in $T_{c}$ correspond to couplings where the impurity band is forming and
separating from the valence band. On the other hand, $T_{c}$ does not change once the impurity band is 
fully separated from the valence band.
The maximum $T_{c}$ we obtained is $\sim$220 K which is closer to experimental 
results~\cite{k_wang_04} than previous estimates using the $k\cdot p$ 
model.~\cite{k_aryanpour_05a,m_majidi_06a} One can also see the same behavior from the inset which 
shows the  average magnetization of the Mn$^{+2}$ ions versus temperature. However, note that within 
the DMFA the average magnetization is not frustrated. This is an artifact of DMFA. Inclusion of 
non-local correlations using cluster methods reduces the value of the magnetization.~\cite{u_yu_10}

Figure~\ref{Tc.vs.nf} displays $T_{c}$ versus hole filling for $x$=5$\%$ and different couplings. 
For large $J_{c}$, in the impurity-band regime, $T_{c}$ is maximum at the optimum filling of $n_{h}=x/2$.
This is consistent with studies based on the impurity band picture.~\cite{k_aryanpour_05a,j_moreno_06a} 
For moderate couplings around $J_{c}\sim$1.5 eV and fillings $n_{h} >$ 2\%, $T_{c}$ increases by decreasing 
the hole concentration. Recent studies in thin films find $T_{c}$ to be proportional to the 
hole concentration for a wide range of $n_{h}$ values.\cite{m_sawicki_10a,g_acbas_09,y_nishitani_10a} 
This might indicate a filling larger than the doping, $n_{h} >x$,  in those experiments.
The inset in  Figure~\ref{Tc.vs.nf} shows $T_{c}$ versus $J_{c}$ for fixed hole concentration 
$n_{h}=x/2=2.5\%$. The plateau at large couplings is due to the absence of non-local correlations 
within the DMFA.

\begin{figure}[t]
\begin{centering}
\includegraphics[width=0.45\textwidth]{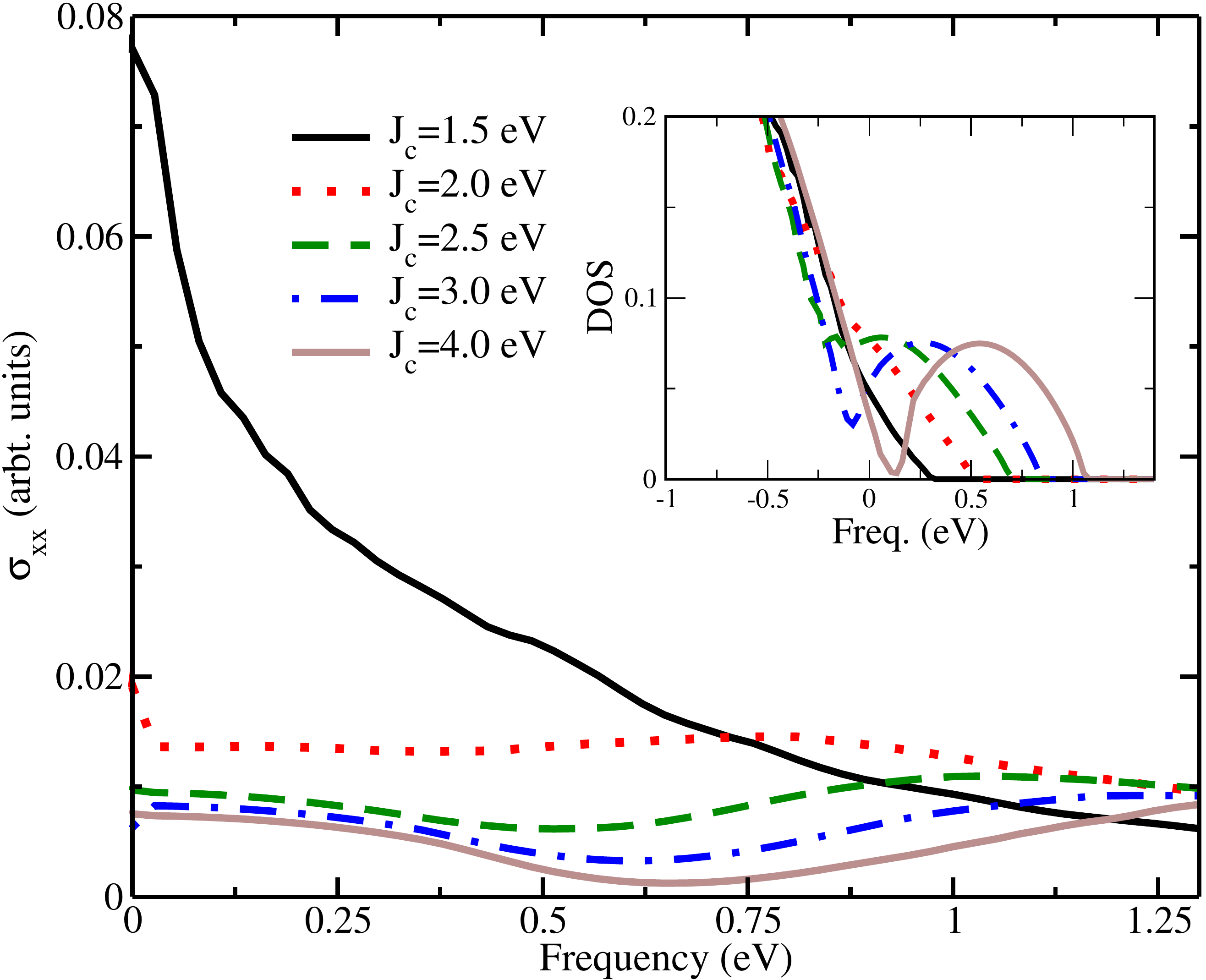}
\caption{(color online). Optical conductivity in the direction perpendicular to the magnetization 
for different couplings at $x$=5\%, $n_{h}$=2.5$\%$ and $T$=58 K. The conductivity is suppressed by 
increasing of $J_{c}$ due to the increasing localization of the holes. The inset displays the total 
density of states versus energy at the edge of the valence band for the same parameters.}
\label{conduct}
\end{centering}
\end{figure}

Finally, we calculate the optical conductivity 
in the direction perpendicular to the magnetization, $\sigma_{\text{xx}}$, 
for energies smaller than the band gap and different couplings (Fig.~\ref{conduct}). 
For small couplings, the conductivity displays a Drude like peak which 
is suppressed by increasing the coupling. This suppression is due to the 
increasing bounding between the magnetic ions and the itinerant holes.
Our results are in agreement with other theoretical calculations.\cite{m_turek_09} 
However, we do not capture the low energy ($\sim$0.2 eV) peak 
observed in  measurements of the conductivity.\cite{k_burch_06} 
The origin of this peak is still highly controversial. While Burch {\it et al.} explain it as  
evidence of the existence of the impurity band\cite{k_burch_06}, others explain the result within 
the distorted valence band picture.\cite{g_acbas_09} 
Since our results for the density of states do not show any low energy feature for $J_{c}\leq$2.0 eV, 
we do not expect to see any peak in optical conductivity for $J_{c}\leq$2.0 eV. 
Moreover, for large couplings the wide impurity band will induce a feature with similar width.
The inclusion of non-local correlations reduces the bandwidth of the impurity band and, it might
be possible to capture this low energy peak within a dynamical cluster study.~\cite{u_yu_10}

\paragraph*{Conclusions-}
We study  the magnetic and transport properties of the diluted magnetic 
semiconductor Ga$_{1-x}$Mn$_{x}$As within the framework of the multi-orbital DMFA. 
We employ a semi-empirical $sp^{3}$ tight-binding approximation to model the band 
structure of the parent material. We choose
tight-binding parameters according to Chadi,\cite{d_chadi_77} which reproduce  the correct band 
structure within the relevant energy window,  including the  
valence, split-off and conduction bands. The spin-orbit coupling is modeled with a  term 
$\lambda_{\alpha} {\bf L} \cdot {\bf S}$. We find that this more realistic band structure leads to 
more realistic results. 

The  density of state shows a distorted valence band for moderate coupling strengths, $J_{c}\sim$1.5 eV, 
and an impurity band for couplings larger than  $J_{c}\sim$3.0 eV.
The band repulsion between the impurity and conduction bands reduces the bandwidth of the impurity 
band. This in turn leads to localization of the itinerant holes and reduction of $T_{c}$. 
The average hole polarization displays a clear magnetic frustration.
Moreover, the  optical conductivity $\sigma_{\text{xx}}$ shows a Drude-like peak at low 
frequencies which disappears with increasing couplings. While our conductivity results are similar
to previous tight-binding calculations,\cite{m_turek_09}  they 
do not display the low energy peak observed experimentally.\cite{k_burch_06} 
Despite the inclusion of band repulsion effects and that our impurity band is narrower than found 
previously, it is still too wide.
This wide impurity band smears out any low energy features. We believe that the inclusion of 
corrections including cavity field and 
non-local correlations into the dynamical mean-field approximation will result in further 
suppression of the impurity bandwidth and, possibly, development of the peak in the conductivity. 

We acknowledge useful conversation with J. Akimitsu, R.~S. Fishman, N. Furukawa, P. Kent, K.  Kubo,
F. Matsukura, H. Ohno, S. Ohya, and M. Tanaka.
This work was supported by the National Science Foundation through
OISE-0952300 and DMR-0548011. Portions of this research were conducted with high performance
computational resources provided by the Louisiana Optical Network Initiative (http://www.loni.org).

\end{document}